\documentstyle{article}
\begin{document}
\newcommand{\beq}{\begin{equation}}
\newcommand{\eeq}{\end{equation}}
\newcommand{\beqn}{\begin{eqnarray}}
\newcommand{\eeqn}{\end{eqnarray}}
\newcommand{\dpf}{\displaystyle\frac}
\newcommand{\no}{\nonumber}
\newcommand{\ep}{\epsilon}
\begin{center}
{\Large Holography and the generalized second law of thermodynamics  in
(2+1)-dimensional cosmology}
\end{center}
\vspace{1ex}
\centerline{\large Bin
Wang$^{a,b,}$\footnote[1]{e-mail:binwang@fma.if.usp.br},
\ Elcio Abdalla$^{a,}$\footnote[2]{e-mail:eabdalla@fma.if.usp.br}}
\begin{center}
{$^{a}$ Instituto De Fisica, Universidade De Sao Paulo, C.P.66.318, CEP
05315-970, Sao Paulo, Brazil \\
$^{b}$ Department of Physics, Shanghai Teachers' University, P. R. China}
\end{center}
\vspace{6ex}
\begin{abstract}
The Fischler-Susskind entropy bound has been studied in (2+1)-dimensional
universes with negative cosmological constant. As in all contracting
universes, that bound is not satisfied. Furthermore, we found that the
Fischler-Susskind bound is not compatible with a generalized second law
of thermodynamics in (2+1)-dimensional cosmology, neither the classical
nor the quantum version. On the other hand, the
Hubble entropy bound has been
constructed in (2+1)-dimensional cosmology and it is shown compatible with
the generalized second law of thermodynamics.
\end{abstract}
\vspace{6ex} \hspace*{0mm} PACS number(s): 04.70.Dy, 98.80.Cq
\vfill
\newpage
Motivated by the well-known result in black hole theory that the total
entropy of matter inside  a black hole cannot exceed the
Bekenstein-Hawking entropy, a  conceptual change in our
thinking about gravity has recently been put forward by the so called
``holographic principle" [1,2]. According to this principle, all the degrees
of freedom inside a volume is expressed on its boundary, implying
that
the entropy of a system cannot be larger than its boundary area. A
specific generalization of the holographic principle to cosmology was
realized by Fischler and Susskind (FS) [3]. A remarkable  point of their
proposal is that the holographic principle is  
valid for  flat or open universes with the equation of state satisfying
the condition $0\leq P\leq\rho$. However, for closed universes the  
principle is violated. The problem becomes even more serious if one 
investigates the universe with a negative cosmological constant [4]. In
that case the holographic principle fails, independently of whether the   
universe is closed, open or flat. Various different modifications of the
FS version of the holographic principle have been raised
recently, such as replacing the holographic principle by the generalized
second law of thermodynamics [5,4], using the cosmological apparent
horizon
instead of the particle horizon in the formulation of holographic
principle  
[6], changing the definition of ``degrees of freedom" [7] etc. A very
recent result claimed that the holographic principle in a closed universe
can be obeyed if the universe contains strange negative pressure matter
[8]. The study of the cosmic holography has also been extended to
Pre-big-bang string cosmological models [9]. All these studies 
have concentrated on (3+1)-dimensional (4D) cosmology.

In our previous work, we have considered the investigations on cosmic
holography in (2+1)-dimensional (3D) cosmological models [10]. Analogously
to the 4D counterpart, the holographic principle is satisfied in all 3D
flat and open universes, but breaks down for 3D closed universes.
Attempts to uphold the holographic principle by introducing negative
pressure matter as well as matter with very unconventional high pressure
failed, because they cannot accomodate any classical description
after the big bang. It is of interest to generalize our
discussions to 3D universes with a negative cosmological constant. There
has been many successful applications of the holographic principle for 3D
pure Anti-de Sitter (AdS) space from string theory [11-14]. Thus we have
the motivation to investigate whether the holographic principle holds in
3D AdS cosmology.

Recently a generalized second law (GSL) of thermodynamics in 4D
cosmologies has been put forward [15], and its relation to the Hubble
entropy bound (HE) suggested by Veneziano [16] has also been discussed.
Further study of the GSL in 4D string cosmology has been addressed as well
[17]. The second purpose of the present paper is to consider their
discussions in 3D cosmological models. By establishing the GSL in 3D
universes, and studying its relation to FS entropy bound, we find that
the FS
bound is not compatible with neither the classical nor the
quantum
mechanical version of GSL. Since the second law of thermodynamics is more
fundamental, the incompatible result between FS bound and GSL
gives us additional reason to look for a reformulation of the cosmological
holographic principle. 

The conflict result between the FS bound and the GSL can be attributed to
the
fact that the FS
bound is too strong and a weaker entropy bound in cosmology is called for.
In 4D universes it was claimed that the HE bound, which is looser than
the FS
bound, is sufficient to avoid any problem with entropy produced at
reheating after inflation [16]. A generalization of these studies to 3D
universes is appealing and will be carried out in this paper. We are going
to define the HE bound and discuss its relation with the FS bound and the
Bekenstein entropy (BE) bound in 3D cosmological models. The relation
between HE bound and GSL will also be addressed and the compatible result
will be reached. These results support the argument that the HE bound
may be a
candidate to replace the FS bound and describe the cosmic holography.

Cosmological solutions in (2+1)-dimensional Einstein gravity have been
proposed in [18,19]. In terms of the (2+1)-dimensional Robertson-Walker
line element
\beq     
{\rm d}s^2={\rm d}t^2-a^2(t)(\dpf{{\rm d}r^2}{1-kr^2}+r^2{\rm d}\theta^2),
\eeq
the Einstein field equations become
\beqn  
(\dpf{\dot{a}}{a})^2+\dpf{k}{a^2} & = & 2\pi G\rho, \\
\dpf{\ddot{a}}{a} & = & -2\pi GP, \\
\dpf{d}{dt}(\rho a^2)+P\dpf{d}{dt}a^2 & = & 0.
\eeqn
When the material content is a perfect fluid with equation of state
\beq 
P=(\gamma -1)\rho,
\eeq
where $\gamma$ is a constant, we derive the relation
\beq 
\rho a^{2\gamma}=const=\rho_0 a_0 ^{2\gamma}.  
\eeq
The scale factor is determined by a Friedmann-like equation
\beq 
(\dpf{\dot{a}}{a})^2=\dpf{2GM_0}{a^{2\gamma}}-\dpf{k}{a^2},
\eeq
where $M_0=\pi\rho_0 a_0 ^{2\gamma}$. For $\gamma=1$, the universe is  
dust-filled and always expands regardless of the value of $k$. However,   
for
$1<\gamma\leq 2$, the solutions of (7) are closed, open or flat
cosmological models according to whether $k$ is $1, -1,$ or $0$,
respectively. The case $\gamma=3/2$ corresponds to the radiation-dominated
universe.  

It is of interest
to investigate the holographic principle in a universe with negative
vacuum energy. A universe with negative cosmological constant is
contracting, independently of the value of $k$.
For simplicity, we just consider the flat universe ($k=0$) with
general equation of state ($1<\gamma\leq 2$). The vacuum
energy density is negative, $-\lambda<0$, so that in the expanding
universe
$\rho '=\rho -\lambda =\dpf{\rho_0 a_0 ^{2\gamma}}{a^{2\gamma}}-\lambda$,
and the Friedmann
equation can be
expressed as
\beq
\dot{a}^2=\dpf{2GM_0}{a^{2\gamma -2}}-\lambda a^2.
\eeq
The scale factor can be calculated and has the form
\beq  
a(t)=(2GM_0/\lambda)^{1/(2\gamma)}\{\sin[\gamma\sqrt{\lambda}t]\}
^{1/\gamma}. 
\eeq

As in the case discussed in [4,10], we find that $\dot{a}$
vanishes at $a=(2GM_0/\lambda)^{1/(2\gamma)}$, and after that point
$\dot{a}$
becomes negative and the universe collapses. This happens within a finite
time after the beginning of the expansion. One can find the value of $L_H$
at the turning
point:
\beq
L_H(turning)=\dpf{1}{2\gamma\lambda^{1/2}}B((\gamma-1)/2\gamma, 1/2)
\eeq
where $B(p,q)$ is the Euler Beta function. It is worth noting that the
particle horizon here has the same dependence on $\lambda$ as that in
4D case [4]. Putting these formulas together, we get at
the
turning point
\beq
\dpf{S}{A}\sim \lambda^{1/\gamma - 1/2}.
\eeq
Considering $\gamma\leq 2$, for small value of $\lambda$ (usually
believed to be
smaller than $10^{-122}$), the entropy over area bound is satisfied at the
turning point. Now we can consider what happens near the final stage of
collapse, where the universe shrinks to the Planck scale. By symmetry,
$L_H\sim \dpf{2a_0}{a(turning)} L_H (turning)\sim \lambda^{1/2\gamma
-1/2}$ [4] at this time. The scale factor
at the Planck time $t=1$ is
$a(t=1)=(2GM_0/\lambda)^{1/2\gamma}\{\sin[\gamma\sqrt{\lambda}]\}   
^{1/\gamma}$. The ratio is $S/A\sim
\lambda^{1/2\gamma -1/2}$. Hence for small $\lambda$, the ratio
is much bigger than unity for the universe with general equation
of state $1<\gamma\leq 2$. We find that prior to the point
of maximal expansion, the holographic constraints hold. However once
after the point where the ratio exceeds unity the
holographic
bound cannot be further maintained.
This result is in agreement with that in (3+1)-dimensions.

For comparison, we can easily see that in the general d-dimensional case, eq.(4) gets
modified to $\dpf{d}{dt}(\rho a^{d-1})+P\dpf{d}{dt}(a^{d-1})=0$, which in view of the
equation of state (5) has a solution $\rho\sim a^{-\gamma(d-1)}$. Going through
calculations, we learn that Eq(11) has the same form, leading above to similar
conclusions. From this point however, we rather stay in 2+1-dimensions where we are able
to draw further conclusions.

As pointed out in [4], the problem can not be expected to be cured by
replacing particle horizon by the apparent horizon proposed in [6],
because at the turning point $\dot{a}=0$, and Eq.(16) from ref.[6]
diverges, that is,
\beq
\dpf{4\sigma}{3a^2 (t)\dot{a}(t)}\sim \infty.
\eeq
Therefore the new holographic principle is violated even earlier.

We cannot naively expect the bound to be saved by considering  
the universe with the unusual negative pressure matter analoguous to the
4D closed universe case [8], because as we see
from
(8),  after the big bang the universe cannot have sufficient
expansion for $\gamma <1$, therefore the classical description is not
valid.

It is appealing to establish the GSL in 3D universes and study the
relation between the GSL and the FS bound. Using the idea proposed in
[16], the
definition of the total entropy of a domain in 3D universes containing
more than one cosmological horizons is given by $S=n_H S^H$, where $n_H$
is a number of cosmological horizons within a given comoving ``volume"
divided by a ``volume" of a single horizon $n_H=a(t)^2/\vert
H(t)\vert^{-2}$; $S^H$ is the entropy within a given horizon. The 
classical
GSL requires that the cosmological evolution must obey $dS\geq 0$, which
corresponds to 
\beq 
n_H\partial _t S^H+ \partial _t n_H S^H \geq 0
\eeq

Adopting the idea used in the GSL for black hole, we consider that there
could
be many sources and types of entropy, and the total entropy is
the sum of their contributions. Supposing a single type of
entropy is dominant, $S^H=\vert H\vert^\alpha$, where $\alpha$ indicates
the type of the entropy source, therefore $S=(a\vert H\vert)^2\vert
H\vert^\alpha$, and Eq(13) can be rewritten as 
\beq     
2H+(2+\alpha)\dpf{\dot{H}}{H}\geq 0.
\eeq
Let us reexpress Eqs.(2-4) in the forms,
\beqn    
H^2 & = & 2\pi G\rho -\dpf{k}{a^2} \no \\   
\dot{H} & = & -2\pi G (P+\rho)+\dpf{k}{a^2} \\
\dot{\rho} & + & 2H(P+\rho)= 0 \no
\eeqn
and substitute them into (14). We find that the relations for equations of
state determined by the GSL are
\beqn   
\dpf{P}{\rho} & \leq & \dpf{2}{2+\alpha} -1+\dpf{\alpha k}{2\pi
G(2+\alpha)a^2\rho}, \hspace{0.5cm} for \hspace{0.1cm} H>0, \\
\dpf{P}{\rho} & \geq & \dpf{2}{2+\alpha} -1 +\dpf{\alpha k}{2\pi
G(2+\alpha )a^2\rho}, \hspace{0.5cm} for \hspace{0.1cm} H<0.
\eeqn
In the last terms of Eqs(16,17), $a^2\rho$ corresponds to the energy of
the whole universe, so $E=a^2\rho \gg k$, thus the last terms in Eqs.
(16,17) can be neglected.

Employing the first law of thermodynamics, $TdS=dE+PdV=(\rho
+P)dV+Vd\rho$,
the
temperature can be obtained by $T^{-1}=(\dpf{\partial S}{\partial
E})_V=\dpf{\partial s}{\partial \rho}$, where $E=\rho V, S=sV$, 
therefore
\beq    
T=\dpf{1}{\pi G(2+\alpha)\vert H\vert^\alpha}
\eeq

To ensure that singularities are avoided for the expressions of the total
entropy $S$, $\partial _t S$ and the temperature $T$ when a flat space
limit
of vanishing $H$ is taken into account, the reasonable physical range of
$\alpha$ should lie within the region $-1\leq \alpha \leq 0$. 

Let us now consider that the dominant contribution to the entropy of the
universe is given by
the geometric entropy $S_g$ whose source is the existence of a
cosmological horizon [20,21]. This is a speculative notion introduced in [15]. Let us
herewith suppose that a component of entropy arises from geometry.  We are thus in a
position
to discuss the relation between the GSL and the FS entropy bound. For a
system
with a cosmological horizon, $S_g ^H$ is given by 
\beq   
S_g ^H=\vert H\vert ^{-1}G_N ^{-1},
\eeq
which corresponds to $\alpha =-1$. Substituting this value of $\alpha$ into
Eqs.(16,17), equations of state corresponding to adiabatic evolution
with dominant $S_g$ are obtained. For the expanding universe $H>0$, and GSL
requires $P\leq \rho$, which is in agreement with the results obtained in
[10] for the FS bound. However for negative $H$, which corresponding to
the
contracting universe, GSL requires $P/\rho \in (1, \infty)$. This range
corresponds to $\gamma >2$ in Eq.(5), which is ruled out in any 3D
contracting universes in FS bound discussions. Therefore, for the 3D
contracting
universes the FS bound is not compatible with the GSL. 

Whether adding a missing
quantum entropy term and developing the quantum mechanical version of the  
GSL
can help us arriving at the compatibility between the FS bound and the GSL
for
contracting 3D universes is still not clear. Using the definition for the
quantum entropy in 4D cases [15], $dS_{Quan.}=-\mu dn_H$, where $\mu$ is a
``chemical potential",  always taken to be positive, $n_H=(aH)^2$, we obtain
\beqn 
dS & = & dS_{Class.}+dS_{Quan.} \no \\
   & = & dn_H S^H+n_H dS^H-\mu dn_H
\eeqn
where $S^H$ is the classical entropy within a cosmological horizon and
$S^H=\vert H\vert^\alpha$ if the classical entropy is dominated by a
single source. The quantum modified GSL can
be expressed as
\beq   
(2H+2\dpf{\dot{H}}{H})n_H(S^H -\mu)+\alpha\dpf{\dot{H}}{H}n_H S^H \geq 0.
\eeq
Considering that geometric entropy still dominates the classical
entropy, $\alpha=-1$, we learn from (21) that
\beq 
\dpf{P}{\rho}\geq\dpf{S^H}{S^H -2\mu}
\eeq
for the contracting universe, $H<0$.

If $\mu\ll S^H$, it returns to the classical case and leads to
$\dpf{P}{\rho}\in (1, \infty)$; if quantum effects are comparable to
classical effects, $\mu \sim S^H$, $\dpf{P}{\rho}\in (-1, \infty)$; and if
quantum effect dominates, say $\mu\gg S^H$, $\dpf{P}{\rho}\in (0,
\infty)$.
These results are never compatible with the requirement of the FS bound,
because $\dpf{P}{\rho}>1$ or $-1<\dpf{P}{\rho}<0$ corresponds to $\gamma
>2$ or $ 0<\gamma<1$, respectively, which are all unacceptable by the FS
bound in all
contracting universes. Therefore the quantum consideration still cannot
lead to the compatibility between the FS bound and the GSL.

Since we believed that the second law of thermodynamics is a
fundamental principle in physics, such an incompatible result gives us
additional motivation to seek for the reformulation of the cosmic
holographic
principle. One may attribute the incompatibility to the argument that FS
bound is too strong, a looser cosmic entropy bound is called for to solve
this conflict with GSL. A possible way was suggested by Veneziano for 4D
cosmology [16] by replacing the FS bound by the HE bound. Based upon the
argument that a black hole larger than $H^{-1}$ cannot form, generalizing
to 3D cases we find the largest entropy in a region corresponding to have
just one black hole per Hubble ``volume" $H^{-2}$ is that $s\leq M_p
^2\vert H\vert$, where $M_p=G_N^{-1/2}$, and the HE bound in 3D universes
can be defined as
\beq    
S^H\leq M_p ^2\vert H\vert^{-1}\sim S_{HE}
\eeq
Recalling the definitions for the BE bound and the FS bound, in 3D they
can be
expressed as
\beqn       
S_{BE} & \sim & EL/\hbar\sim Md_p/\hbar\approx \rho d_p^2\cdot
d_p/\hbar\sim H^2 d_p^2\cdot d_p/l_p^2 \\
S_{FS} & \sim & d_p/\hbar\sim d_p/l_p^2
\eeqn
where $d_p$ is the particle horizon, $l_p\sim\sqrt{\hbar}\sim M_p$. We 
substitute the particle horizon $d_p$ by $H^{-1}$, and consider that
when applied to non-inflationary cosmology, they are about the same [22].
Thus in 3D cases we can reproduce the relation among different entropy
bounds
first obtained in 4D cases [16],
\beq   
S_{HE}=S_{FS}^{1/2} S_{BE}^{1/2}
\eeq
It is easy to see that the HE bound is much looser than the FS bound. 

The most attractive point now is to study whether the HE bound is
compatible with GSL in 3D cosmological models. Assuming a single dominant
entropy form, $S^H=(\dpf{\vert H\vert}{M_p})^{\alpha}$, bound (23) can be
written as 
\beq        
(\dpf{\vert H\vert }{M_p})^\alpha\leq M_p^2\vert H\vert^{-1}
\eeq
Considering the reasonable physical region of $\alpha$, we need
\beq       
\vert H\vert\leq M_p\cdot M_p^{1/\alpha+1}.
\eeq

From the HE bound, we learn that $\vert H\vert$ has a maximum value,
therefore in order not to violate this bound we need for expanding
universe $H>0$, the evolution of the 3D universe undergoes decelerated
expansion, say $H>0, \dot{H}<0$. GSL allows such an evolution. Since in the
physical region $-1\leq\alpha\leq 0$, Eq(14) reads 
\beq    
2H+(2+\alpha)\dpf{\dot{H}}{H}>2H+\dpf{\dot{H}}{H}\geq 0
\eeq
For $\dot{H}<0$, it leads directly to a requirement for the equation
of state, that is,
 $\rho\geq P$. This is quite natural and thus the entropy bound is
valid without violation of GSL.

For a contracting universe with $H<0$, to satisfy the entropy bound, we
need the universe to experience decelerated contraction, namely
$\dot{H}>0$. This requirement is obviously not compatible with GSL,
because Eqs(14)and (29) tell us that
\beq  
2\vert H\vert^2\leq -\dot{H}
\eeq
which is false for $\dot{H}>0$.

Luckily this conflict between HE bound and GSL can be resolved by
considering the quantum modified GSL (21). For contracting 3D universes,
$H<0$, and Eq(21) can be rewritten as 
\beq    
2\vert H\vert^2n_H(S^H-\mu)\leq -\dot{H}[2n_H S^H-2n_H\mu+\alpha n_H S^H].
\eeq
Neglecting the quantum effect, $\mu\ll S^H$, Eq(31) boils down to (30) for
$\alpha =-1$. However if we consider that the quantum effect is
strong,
namely $S^H -\mu<0$, we can rewrite Eq(31) as
\beq   
2\vert H\vert^2\geq -\dot{H}\mid 2+\dpf{\alpha S^H}{S^H-\mu}\mid.
\eeq
For decelerated contraction $\dot{H}>0$, Eq(32) certainly holds.
Thus considering the quantum effect, the conflict between HE bound and GSL
can be overcome.

In summary we have found that analogously to 3D closed universes, FS bound
in 3D
cosmological models with negative cosmological constant breaks down
regardless of the value of $k$. Unlike the 4D closed universe [8], here the negative
pressure
matter cannot be used to save the FS bound. Establishing the GSL in 3D cosmologies, we
have shown that the state equations required are not consistent with
those needed by FS bound, which shows that FS bound and GSL are not
compatible. Furthermore we have shown that the conflict cannot be resolved by taking
account of the
quantum modified version of GSL, which has not been addressed in 4D cases. Considering
that the GSL is a fundamental
principle in physics, we have further motivations to rethink the
expression of
cosmic holography. Extending different cosmic entropy bounds to 3D
cosmology, we have reproduced the relation between HE bound, BE bound and
FS bound first obtained in 4D cases [16]. Compared to FS bound, HE bound
is looser, and compatible with GSL. This result agrees with that
claimed in 4D cosmology [15] and supports the argument that HE bound is a
candidate for describing cosmic holography. It is of interest to consider generalization
of our discussions on the relation between GSL and cosmic entropy bound to higher
dimensional spacetimes.

ACKNOWLEDGEMENT: This work was partically supported by
Fundac$\tilde{a}$o de Amparo $\grave{a}$ Pesquisa do Estado de
S$\tilde{a}$o Paulo (FAPESP) and Conselho Nacional de Desenvolvimento 
Cient$\acute{i}$fico e Tecnol$\acute{o}$gico (CNPQ).  B. Wang would also
like to acknowledge the support given by Shanghai Science and Technology
Commission.

\end{document}